\documentclass[useAMS,usenatbib]{mn2e}
\usepackage{graphicx}
\usepackage{lscape}
\usepackage{rotating}
\newcommand{\Msun}{\mbox{M$_{\odot}$}}
 
\begin{document}
%
%
%
%


\def\aj{\rm{AJ}}                   
\def\araa{\rm{ARA\&A}}             
\def\apj{\rm{ApJ}}                 
\def\apjl{\rm{ApJ}}                
\def\apjs{\rm{ApJS}}               
\def\ao{\rm{Appl.~Opt.}}           
\def\apss{\rm{Ap\&SS}}             
\def\aap{\rm{A\&A}}                
\def\aapr{\rm{A\&A~Rev.}}          
\def\aaps{\rm{A\&AS}}              
\def\azh{\rm{AZh}}                 
\def\baas{\rm{BAAS}}               
\def\jrasc{\rm{JRASC}}             
\def\memras{\rm{MmRAS}}            
\def\mnras{\rm{MNRAS}}             
\def\pra{\rm{Phys.~Rev.~A}}        
\def\prb{\rm{Phys.~Rev.~B}}        
\def\prc{\rm{Phys.~Rev.~C}}        
\def\prd{\rm{Phys.~Rev.~D}}        
\def\pre{\rm{Phys.~Rev.~E}}        
\def\prl{\rm{Phys.~Rev.~Lett.}}    
\def\pasp{\rm{PASP}}               
\def\pasj{\rm{PASJ}}               
\def\qjras{\rm{QJRAS}}             
\def\skytel{\rm{S\&T}}             
\def\solphys{\rm{Sol.~Phys.}}      
\def\sovast{\rm{Soviet~Ast.}}      
\def\ssr{\rm{Space~Sci.~Rev.}}     
\def\zap{\rm{ZAp}}                 
\def\nat{\rm{Nature}}              
\def\iaucirc{\rm{IAU~Circ.}}       
\def\aplett{\rm{Astrophys.~Lett.}} 
\def\apspr{\rm{Astrophys.~Space~Phys.~Res.}}
\def\bain{\rm{Bull.~Astron.~Inst.~Netherlands}} 
\def\fcp{\rm{Fund.~Cosmic~Phys.}}  
\def\gca{\rm{Geochim.~Cosmochim.~Acta}}   
\def\grl{\rm{Geophys.~Res.~Lett.}} 
\def\jcp{\rm{J.~Chem.~Phys.}}      
\def\jgr{\rm{J.~Geophys.~Res.}}    
\def\jqsrt{\rm{J.~Quant.~Spec.~Radiat.~Transf.}}
\def\memsai{\rm{Mem.~Soc.~Astron.~Italiana}}
\def\nphysa{\rm{Nucl.~Phys.~A}}   
\def\physrep{\rm{Phys.~Rep.}}   
\def\physscr{\rm{Phys.~Scr}}   
\def\planss{\rm{Planet.~Space~Sci.}}   
\def\procspie{\rm{Proc.~SPIE}}   

\let\astap=\aap
\let\apjlett=\apjl
\let\apjsupp=\apjs
\let\applopt=\ao

\title[Blue not brown: UKIDSS T dwarfs with suppressed $K$-band fluxes]{Blue not brown: UKIDSS T dwarfs with suppressed $K$-band flux}

\author[D.N. Murray et al.]{D.~N.~Murray$^{1}$\thanks{E-mail: D.Murray@herts.ac.uk}, 
B.~Burningham$^{1}$, H.~R.~A.~Jones$^{1}$, D.~J.~Pinfield$^{1}$,
\newauthor 
P.~W.~Lucas$^{1}$,
S.~K.~Leggett$^{2}$,
C.~G.~Tinney$^{3}$,
A.~C.~Day-Jones$^{4}$, 
D.~J.~Weights$^{1}$,
\newauthor
N.~Lodieu$^{5}$,
J. A. P\'erez Prieto$^{6}$,
E.~Nickson$^{7}$,
Z.~H.~Zhang$^{1}$,
J.~R.~A.~Clarke$^{1}$,
\newauthor
J.~S.~Jenkins$^{4}$,
M.~Tamura$^{8}$
\\
$^{1}$ Centre for Astrophysics Research, Science and Technology Research Institute, University of Hertfordshire, Hatfield AL10 9AB;\\
$^{2}$ Gemini Observatory, 670 N. A'ohoku Place, Hilo, HI 96720, USA;\\
$^{3}$ School of Physics, University of New South Wales, 2052, Australia;\\
$^{4}$ Departmento de Astronomia, Universidad de Chile, Santiago, Chile;\\
$^{5}$ Instituto de Astrof\'isica de Canarias (IAC), Calle V\'ia L\'actea s/n, E-38200 La Laguna, Tenerife, Spain;\\
$^{6}$ Departamento de Astrof\'isica, Universidad de La Laguna (ULL), E-38205 La Laguna, Tenerife, Spain;\\
$^{7}$ School of Physics and Astronomy, University of Southampton, SO17 1BJ;\\
$^{8}$ National Astronomical Observatory, Mitaka, Tokyo, 181-8588\\
}

\date{28th March 2010}

\pagerange{\pageref{firstpage}--\pageref{lastpage}} \pubyear{2010}

\maketitle

\label{firstpage}

\begin{abstract}
We have used blue near-infrared colours to select a group of 12 spectroscopically-confirmed UKIDSS T dwarfs later than T4. From amongst these we identify the first two kinematic halo T-dwarf candidates.
 Blue near-infrared colours have been attributed to collisionally-induced hydrogen absorption, which is enhanced by either high surface gravity or low metallicity.
Proper motions are measured and distances estimated, allowing the determination of tangential velocities.
$U$ and $V$ components are estimated for our objects by assuming $V_{rad}=0$. From this, ULAS~J0926+0835 is found to have $U=62$ km\,s$^{-1}$ and $V=-140$ km\,s$^{-1}$ and ULAS~J1319+1209 is found to have $U=192$ km\,s$^{-1}$ and $V=-92$ km\,s$^{-1}$.
These values are consistent with potential halo membership. However, these are not the bluest objects in our selection. The bluest is ULAS~J1233+1219, with $J-K=-1.16\pm0.07$, and surprisingly this object is found to have young disc-like $U$ and $V$.
 Our sample also contains Hip~73786B, companion to the metal-poor K5 dwarf Hip 73786.
Hip 73786 is a metal-poor star, with $[Fe/H]=-0.3\pm0.1$ and is located at a distance of 19$\pm0.7$ pc. $U, V, W$ space velocity
 components are calculated for Hip 73786A and B, finding that $U=-48\pm7$ km\,s$^{-1}$, $V=-75\pm4$ km\,s$^{-1}$ and $W=-44\pm8$ km\,s$^{-1}$. From the properties of the primary, Hip~73786B is found to be at least $1.6$ Gyr old.
As a metal poor object, Hip~73786B represents an important addition to the sample of known T dwarf benchmarks. 
\end{abstract}

\begin{keywords}
surveys - stars: low-mass, brown dwarfs
\end{keywords}

\section{Introduction}
\label{sec:Intro}
The first definitive detection of a T dwarf, Gliese 229B, occurred only as recently as 1995 \citep{Nakajima_1995}. Many more have been found in sky surveys such as the Sloan Digital Sky Survey \citep[SDSS; ][]{Abazajian_2003}, the Two-Micron All-Sky Survey  \citep[2MASS;][]{Cutri_2003}, the Canada-France Hawaii Telescope's Brown Dwarf Survey \citep[CFBDS;][]{Delorme_2008} and the UKIRT Infrared Deep Sky Survey \citep[UKIDSS;][]{Warren_2007}. 
Subdwarfs have been known for longer, e.g. \citet{Kuiper_1939}. These are very metal-poor stars (Gizis 1997 suggests $[m/H] = -1.2$ as a typical value for subdwarfs).  Subdwarfs also have fainter absolute magnitudes than solar-abundance
 stars with the same $B-V$ colour \citep{Gizis_1997}.

Along with globular clusters and tidal streams from captured dwarf galaxies, the
 Milky Way's stellar halo also contains many subdwarfs. Subdwarfs represent the remnant members of earlier generations of star formation, when the interstellar medium had experienced less enrichment by supernovae and red giants. \citet{Reid_1998} estimates an age of 11 to 13 Gyr for the halo.
 The substellar extension to the subdwarf luminosity sequence could provide an important test of theories of brown dwarf formation in this environment. 
Obtaining accurate constraints on the substellar luminosity function is an important step toward clearer understanding of the physics underlying brown dwarf and star formation,
and in particular the role played by metallicity. Unlike the disc stars, however, the substellar-mass extension of the halo subdwarfs is not well sampled.

Such objects also provide a unique environment in which to test our understanding of the physics of metal-poor atmospheres. Although the properties of low-metallicity atmospheres have 
been modelled at a variety of temperatures \citep[e.g.,][]{Saumon_1994,Lenzuni_1991}, there are currently few objects with well-established, subsolar metallicites against which the models may be tested.

L subdwarfs are known to exist, two examples of such objects being 2MASS~J05325346+8246465 \citep{Burg_2003} and ULAS J135058.86+081506.8 \citep{Lodieu_2010}. Brown dwarfs cool as they age, evolving through the L and into the T spectral
classes \citep{Burrows_2001} and given the substantial ages of halo stars, there has been ample time for objects that were originally halo L dwarfs to evolve into the T class.  Significantly, the L subdwarfs described above are known to posses halo kinematics.

Several T dwarfs have been suggested as metal-poor.
 Amongst these objects are J12373919+6526148 \citep{Vrba_2004} with $[m/H] \sim -0.2$ \citep{Liebert_2007}, 2MASS~J11145133-2618235 \citep{Tinney_2005} with $[m/H]\sim -0.3$ \citep{Burg_2006},
 2MASS~J09393548-2448279 \citep{Tinney_2005} with $[m/H] \sim -0.3$ \citep{Leggett_2007} and also Epsilon Indi Ba and Bb with $[m/H] \sim -0.2$ \citep{Santos_2004}.
In addition to these low-metallicity objects, there are several other T dwarfs which have been considered as subdwarf candidates.
 These are 2MASS~J0937+2931 \citep{Burg_2002}, SDSS~J1416+1348B \citep[see ][]{Ben_2010,Scholz_2010a} and CFBS~J1500-1824, discovered by Delorme et al. and reported in \citet{Burg_2009}.
SDSS~J1416+1348B, CFBS~J1500-1824 and 2MASS~J0937+2931 all show heavily-depressed $K$-bands, due to enhanced collisionally-induced hydrogen absorption \citep[hereafter, CIA H$_2$; see][]{Linsky_1969}.
 In addition, 2MASS~J0937+2931 and SDSS~J1416+1348B have unusual, broadened $Y$-bands \citep[respectively][]{Burg_2006,Burg_2010}.
2MASS~J0937+2931 shows little evidence of potassium absorption in its $J$-band \citep{McLean_2007}.
CFBS~J1500-1824 also does not show the 1.25$\mu$m potassium doublet, an absence that would be consistent with low metallicity.

However, none of these T dwarfs show unambigously halo-like kinematics. \citet{Vrba_2004} found 2MASS~J0937+2931's tangential velocity to be young disc-like, 47 km\,s$^{-1}$.
 \citet{Bowler_2010} report velocity components for the SDSS~J1416+1348AB system such that $(U, V, W)$ = $(6\pm4, 10.2\pm1.2, -27\pm9)$, which are young disc values.
 The kinematics of CFBDS~J1500-1824 were described as implying an 80 percent probability that it is part of the old disc, with a 10 percent probability of halo membership.

\section{Identifying candidates}
\label{sec:Cands}

The volume probed for T dwarfs by the UKIRT Infrared Deep Sky Survey \citep[UKIDSS; ][]{Lawrence_2007} Large Area Survey (LAS) is much larger than previously available, making feasible the identification of the T subdwarf population.
Many T dwarfs have now been published as part of an ongoing program to spectroscopically confirm many T dwarfs in the UKIDSS LAS, as reported in \citet{Burningham_2010} and references therein, and it is from this group that we have selected the targets investigated in this paper.

\begin{figure*}
\includegraphics[height=500pt, angle=90]{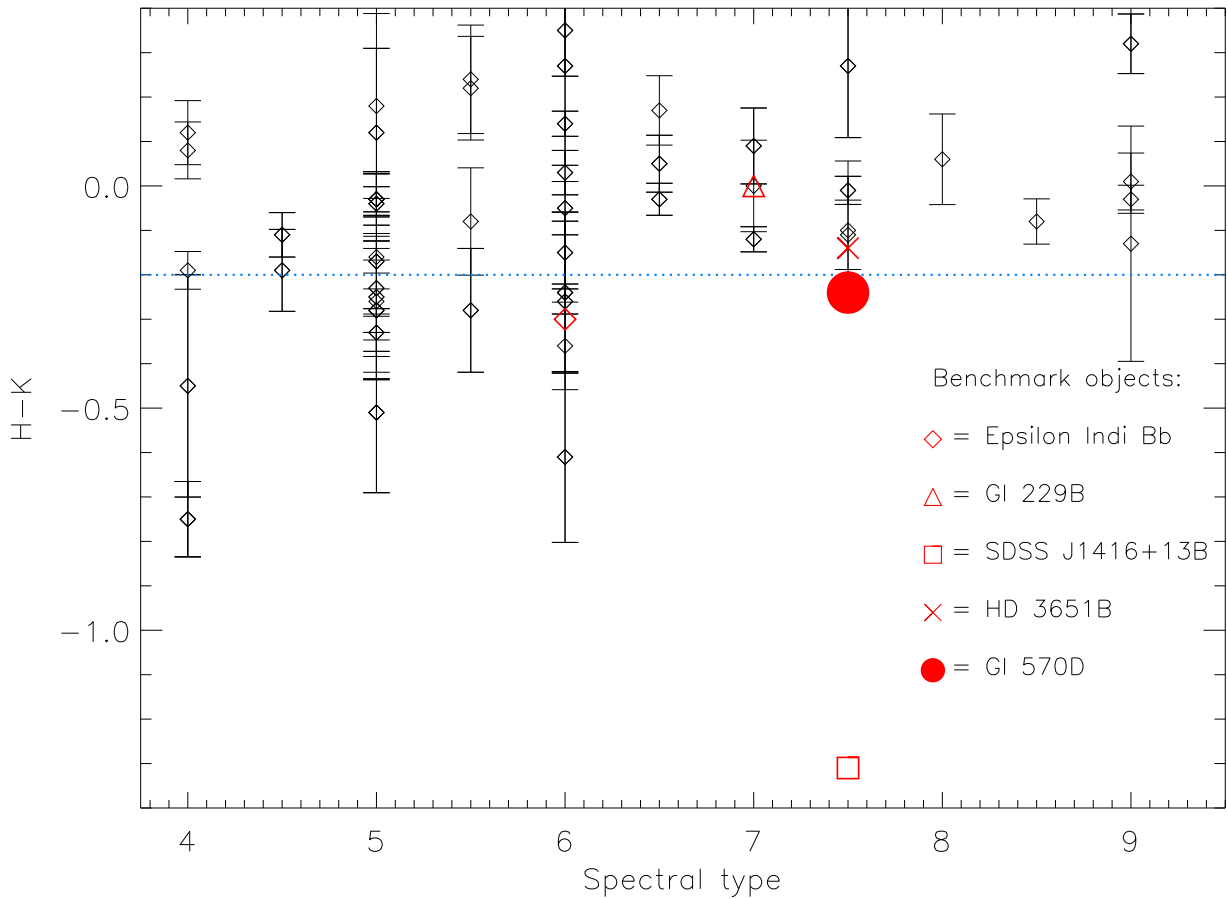}
\caption{A plot showing $H-K$ plotted against spectral type. Black diamonds represent UKIDSS T dwarfs \citep[see ][]{Burningham_2010,Pinfield_2008,Lodieu_2007}, with the error bars showing the uncertainty in $H-K$. The other symbols, in red, represent binary-benchmark T dwarfs.  Also included as a further point of comparison is SDSS~J1416+1348B, from \citet{Ben_2010}.
The blue dashed line represents our $H-K=-0.2$ selection criterion.}
\label{fig:Select}
\end{figure*}

 The UKIDSS sample of T dwarfs were themselves selected through a set of colour cuts, requiring $J-H < 0.1$ and $J-K < 0.1$. Then, after cross-matching with SDSS, an
 object would be required either to have $z-J>3.0$ or to be undetected in in $z$. Potential metal-poor T dwarfs were identified from amongst these spectroscopically-confirmed objects using a $H-K$ vs. spectral type plot (shown in Figure~\ref{fig:Select}).
The UKIDSS sample is most complete for targets later than a spectral type of T4, so a type later than T4 was required along with the $H-K < -0.2$ cut. 

The $H-K$ cut is intended to select blue outliers relative to the population, as the 
 $K$-band is expected to be the most strongly-affected by CIA H$_2$ absorption, and this colour selection should thus prefer a set of objects showing suppressed $K$-band flux.
 Comparisons to T dwarfs from binary systems, whose primary stars' provide fiducial constraints on metallicity, support our criteria. HD 3651B, with an $[Fe/H]=0.12\pm0.04$ \citep{Santos_2004} also lies outside of our selection area. Gliese 570D, with an essentially solar metallicity of $[Fe/H]=0.04\pm0.06$ \citep{Feltzing_1998}, lies on the boundary,
whereas the metal-poor T6 Epsilon Indi Bb, with $[Fe/H]=-0.23\pm0.06$ \citep{Santos_2004}, lies within the selection area. 

We also note that $K$-band flux is sensitive to surface gravity as well as metallicity. The opacity of CIA H$_{2}$ absorption varies with
 the square of the local gas number density \citep{Saumon_1994}. For a given $T_{\rm eff}$, higher-gravity T dwarfs will have higher gas pressures and thus tend to have larger H$_{2}$ opacity. As such they will be fainter in the $K$-band, leading 
to bluer $H-K$ colours than comparable lower-gravity objects. This surface gravity-related CIA H$_{2}$ absorption will be strongest at late-T objects \citep[see][]{Burg_2002,Knapp_2004}.
 In practice, for many objects blue $H-K$ colours may well be due to a combination of gravity and metallicity \citep{Knapp_2004}. This degeneracy means that caution must be exercised before any attempt is made to infer a T dwarf's metallicity
 solely on the basis of its near-infrared colours. For instance, \citet{Liu_2007} show that for a $T_{\rm eff}=900$K object, increasing $log\,g$ from 5.0 to 5.5 dex changes the $K/H$ flux ratio by $\sim$0.25, while
reducing metallicity from $[m/H]=0.0$ to -0.5 changes the ratio by $\sim$0.45. Halo objects, however, may well have far lower $[m/H]$ values
\citep[see][]{Gizis_1997}, so for these objects depression of the $K$ band flux may come to be dominated by metallicity effects. The model trend suggests that the bulk of colour variation may be due to metallicity for halo objects.
 Although surface gravity will certainly play a role, the bulk of radius evolution is expected to occur by ages of $\sim$ 1 Gyr for brown dwarfs \citep{Burrows_2001}. This implies a maximum surface gravity for brown dwarfs. \citet{Saumon_2008} find this to be log $g = 5.366$ in the case of a $T_{\rm eff} = 1380$K and 0.068$\Msun$ brown dwarf.
 This futher suggests that the largest source in variation of $H$ and $K$-band colours in halo T dwarfs is likely to be from metallicity.

In addition to the previously-published DR4 LAS T dwarfs a more recently-identified object from DR5, ULAS~J1504+0538, was also included in our $H-K$ selection. 
Since our spectroscopic confirmation of ULAS~J1504+0538 (see Section~\ref{sec:Spectro}), it has been identified by \citet{Scholz_2010} as a common proper motion binary companion to Hip 73786. As such, from here
 on we will refer to this object as Hip~73786B. Our colour-type selection yielded a total of twelve candidates.
 The sample is shown in Figure~\ref{fig:Select}, with $H-K$ plotted in relation to spectral type. The candidates lie below the dashed line. As can be seen, most of our candidates cluster at 
spectral types $<$ T5.5, with none later than T6.

 \begin{table*}
\begin{tabular}[t]{llllllllll}
\hline
Object          & R.A.         & Dec         & Discovery & $J$ mag        & $Y-J$         & $J-H$          & $H-K$          & $J-K$           & Source \\
\hline
ULAS~J0842+0936 &  08 42 11.68 & 09 36 11.78 & 3         & 18.38$\pm0.02$ & 1.2$\pm0.2$   & -0.46$\pm0.03$ & -0.2$\pm0.2$   & -0.7$\pm0.2$    & 4 \\
ULAS~J0926+0835 &  09 26 05.47 & 08 35 17.00 & 3         & 18.57$\pm0.02$ & 1.3$\pm0.2$   & -0.12$\pm0.02$ & -0.5$\pm0.3$   & -0.6$\pm0.3$    & 4 \\
ULAS~J0958-0039 &  09 58 29.86 & -00 39 32.0 & 1         & 18.95$\pm0.06$ & 0.9$\pm0.2$   & -0.5$\pm0.1$   & -0.3$\pm0.2$   & -0.5$\pm0.1$    & 1 \\
ULAS~J1012+1021 &  10 12 43.54 & 10 21 01.70 & 3         & 16.87$\pm0.01$ & 1.13$\pm0.02$ & -0.34$\pm0.02$ & -0.33$\pm0.05$ & -0.68$\pm0.05$  & 4 \\
ULAS~J1018+0725 &  10 18 21.78 & 07 25 47.10 & 2         & 17.71$\pm0.04$ & 1.19$\pm0.09$ & -0.16$\pm0.08$ & -0.3$\pm0.2$   & -0.41$\pm0.2$   & 2 \\
ULAS~J1233+1219 &  12 33 27.45 & 12 19 52.20 & 3         & 17.87$\pm0.03$ & 1.35$\pm0.07$ & -0.41$\pm0.07$ & -0.75$\pm0.08$ & -1.16$\pm0.07$  & 4 \\
ULAS~J1303+0016 &  13 03 03.54 & 00 16 27.70 & 1         & 19.02$\pm0.03$ & 1.2$\pm0.2$   & -0.47$\pm0.09$ & -0.6$\pm0.2$   & -1.1$\pm0.2$    & 1 \\
ULAS~J1319+1209 &  13 19 43.77 & 12 09 00.20 & 3         & 18.90$\pm0.05$ & 1.49$\pm0.07$ &  0.0$\pm0.2$   & -0.5$\pm0.2$   & -0.5$\pm0.1$    & 4 \\
ULAS~J1320+1029 &  13 20 48.12 & 10 29 10.60 & 3         & 17.82$\pm0.02$ & 1.15$\pm0.06$ & -0.07$\pm0.05$ & -0.3$\pm0.1$   & -0.4$\pm0.13$   & 4 \\
ULAS~J1501+0822 &  15 01 35.33 & 08 22 15.20 & 1         & 18.32$\pm0.02$ & 1.4$\pm0.2$   &  0.02$\pm0.03$ & -0.23$\pm0.06$ & -0.21$\pm0.06$  & 1 \\
Hip~73786B      &  15 04 57.66 & 05 38 00.80 & -         & 16.59$\pm0.02$ & 1.05$\pm0.03$ & -0.46$\pm0.04$ & -0.4$\pm0.1$   & -0.82$\pm0.09$  & 3 \\
ULAS~J2320+1448 &  23 20 35.28 & 14 48 29.80 & -         & 16.79$\pm0.02$ & 1.35$\pm0.03$ & -0.35$\pm0.03$ & -0.26$\pm0.03$ & -0.61$\pm0.03$  & 4 \\
\hline
\end{tabular}
\caption{Summary of the $YJHK$ photometric colours of each object. 
The data is all presented on the MKO filter system. Discovery paper references: 1) \citet{Pinfield_2008}; 2) \citet{Lodieu_2007}; 3) \citet{Burningham_2010}. A '-' indicates a spectrum published for the first time here. A $J$-band spectrum for ULAS~J2320+1448 was published in \citet{Burningham_2010}, however the full $JHK$ spectrum shown in Figure~\ref{fig:spec2} is new data.
 The values after the numbers are the magnitude errors. The Source column indicate data sources. These are: 1) \citet{Pinfield_2008}; 
 2) \citet{Lodieu_2007}; 3) Hip~73786B, which is new data; 4) indicates \citet{Burningham_2010}.
\label{tab:colours}}
 \end{table*}

\section{Near-infrared photometry}
\label{sec:Photo}
The near infrared photometry for our selected objects is summarised in Table~\ref{tab:colours}, the bulk of the which has been published in \citet{Burningham_2010} and references therein, where exposure times and observing conditions may also be found.
 All photometry is presented on the Mauna Kea Observatories (MKO) system \citep{Tokunaga_2002}. The full co-ordinates of all objects are show in Table~\ref{tab:colours}, along with references to their discovery papers.

Our photometry for Hip~73786B is derived from our spectroscopy. 
Hip~73786B was observed using the Wide Field CAMera \citep[WFCAM; ][]{Casali_2007} on UKIRT,  on
12 July 2009(UT), with a seeing of $\sim$1.2 arcseconds. The object was 
imaged in $Y$ and $J$ using a 3-point jitter-pattern with 2x2 microstepping. Each individual exposure was 10 seconds.
 The object was observed in $H$ and $K$ with two sets of 5-point jitter patterns at an individual exposure of 10 seconds, with 2x2 microstepping.
 This led to a total exposure in each band of 400 seconds. The data were processed using the WFCAM pipeline by the Cambridge Astronomical Surveys Unit \citep{Irwin_2004}, and archived at the WFCAM Science Archive \citep{Hambly_2008}.

\section{Spectroscopy}
\label{sec:Spectro}

\subsection{New spectra}
Hip~73786B and ULAS~J2320+1448 represent previously-unpublished spectra. 
The sources for previously published spectroscopy are summarised in Table~\ref{tab:colours}.

ULAS~J2320+1448 was observed on the Near InfraRed Imager and Spectrometer \citep[NIRI; see][]{Hodapp_2003} on the Gemini-North telescope to obtain deeper $J$, $H$ and $K$ band spectra on 22 August 2008, 12 October 2008 and 14 October 2008 (UT) respectively. These data were reduced using standard NIRI IRAF packages. The images were flat fielded, masked for bad pixels and median stacked. A dispersion solution was fitted using the arc spectra.
 The method used was in common with that used in \citet{Burningham_2010}.The target was observed at an airmass of 1.01, with an integration time of 750s. The resulting spectrum has an average resolution of $R \tilde 460 $.

To gain a higher signal-to-noise ratio, the short $J$-band discovery spectrum was combined with the new deep $J$-band spectrum, using a weighted average.
 The $J$, $H$ and $K$-band spectra were then scaled by photometry to place them on a common flux scale before combining them to produce a flux-calibrated $JHK$ spectrum.

Hip~73786B was observed on the InfraRed Camera and Spectrograph
\citep[IRCS;][]{Kobayashi_2000} on the 
Subaru telescope on Mauna Kea to obtain R$\sim100$ $JH$ and $HK$ spectra on the nights of 7$^{th}$ May 2009 and 30$^{th}$ December 2009 respectively.
The data were sky subtracted using generic IRAF tools,
and median stacked. An arc frame was used to fit a dispersion solution. The spectra were then 
extracted and cosmic rays and bad pixels were removed using a sigma-clipping algorithm.

Telluric correction was achieved by dividing each extracted target
spectrum by that of an F4V star, which was observed just
after the target and at a similar airmass. 
Prior to division, hydrogen lines were removed from the standard star
spectrum by interpolating the stellar continuum.
Relative flux calibration was then achieved by multiplying through by a
blackbody spectrum of the appropriate $T_{\rm eff}$.
The $JH$ and $HK$ spectra were then joined using the overlap region between 1.43$\mu$m and 1.63$\mu$m in the $H$-band
 to place the spectra on a common flux scale. The overlap region covers a wide range and includes the $H$-band peak.
 As a test of the merger, spectrophotometric colours were computed from the merged spectrum, and these were found to be entirely in agreement with
 the photometry.

\subsection{Spectral types}
\begin{figure*}
\includegraphics[angle=180]{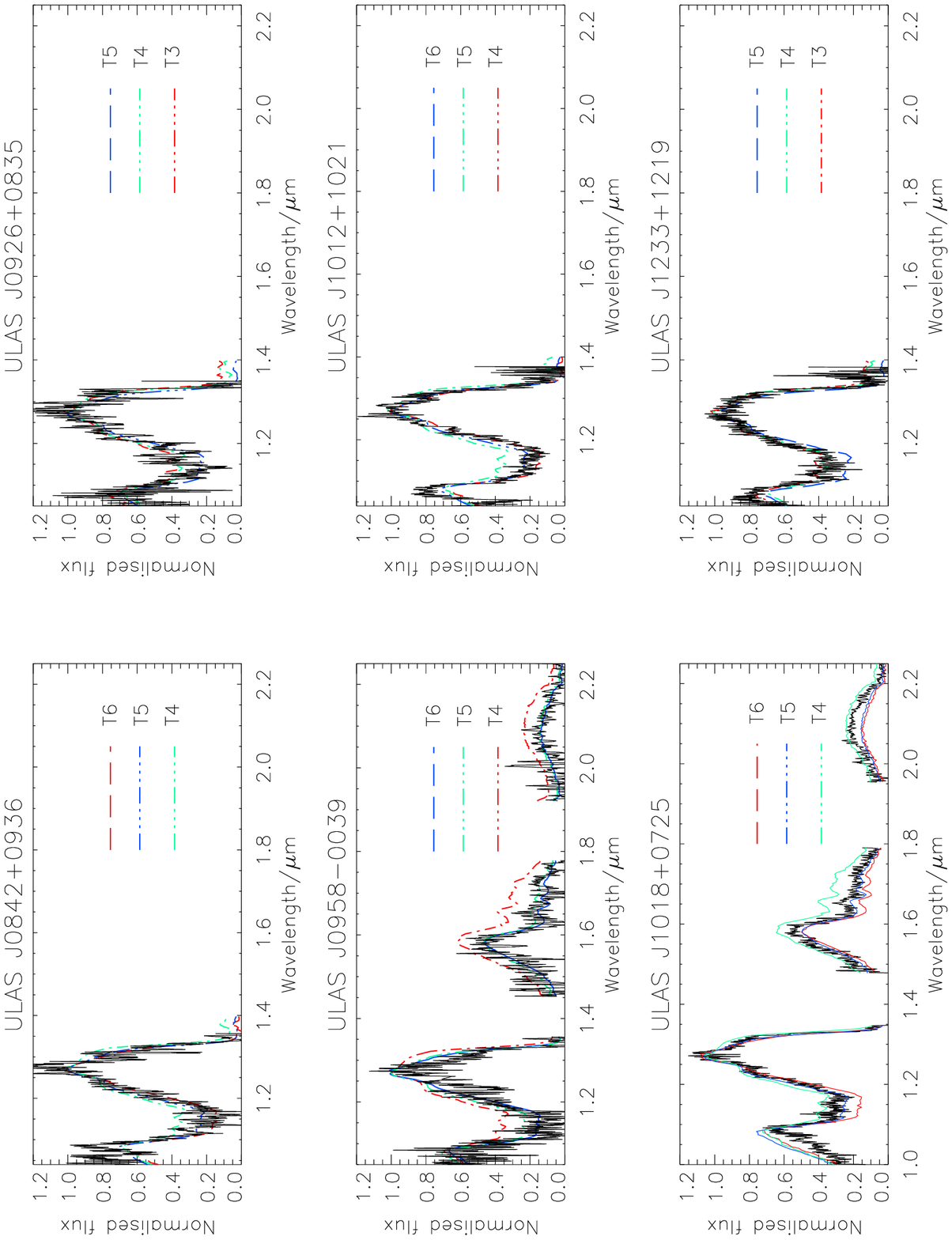}
\caption{Spectra for six candidates, ULAS~J0842+0936, ULAS~J0926+0835, ULAS~J0958-0039, ULAS~J1012+1021, ULAS~J1018+0725 and ULAS~J1233+1219. In each figure, the object spectrum is shown as a solid black line.
 The objects are all shown in comparison
to T dwarf standards. These are represented by the lines on each plot, marked with the respective spectral types. The T3 to T7 spectral standards are from \citet{Burg_2006} and references therein. The object and standard spectra in each plot have been normalised against
 their mean fluxes around 1.27$\mu$m. Each plot has been titled with the name of the object it is for. The gaps in the spectra represent regions where there are no data.}
\label{fig:Sixspectra}
\end{figure*}

\begin{figure*}
\includegraphics[angle=180]{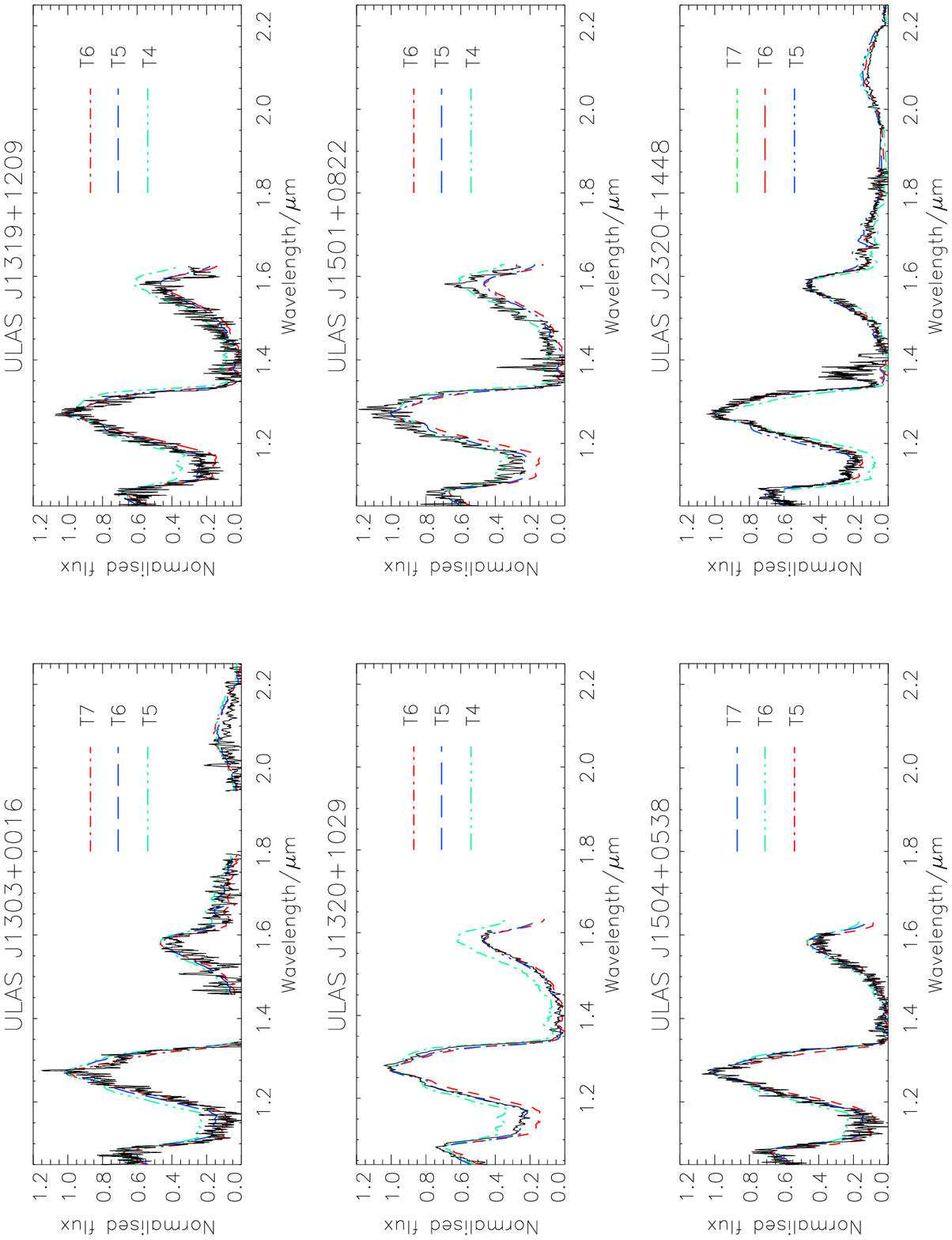}
\caption{Same as Figure 5 for ULAS~J1303+0016, ULAS~J1319+1209, ULAS~J1320+1029, ULAS~J1501+0822, ULAS~J1501+0822, Hip 73786B and ULAS~J2320+1448.}
\label{fig:spec2}
\end{figure*}

Objects were typed using their indices and comparisons to template spectra, following the general procedures 
set out in \citet{Burg_2006}. In brief, these indices target absorption bands of H$_{2}$O and CH$_{4}$ in the $J$-, $H$- and $K$-bands,
 which have been found to correlate with near-infrared spectral type. The template spectra used here are those indicated in \citet{Burg_2006} and references therein.
Typing was conducted through a two-stage process. First, an index-based type was arrived at, using the $J$ and (where available) the $H$-band indices, by taking the median of these values.
 Then, independently, a type was arrived at by plotting the object spectrum against an appropriate range of standard spectra, and visually inspecting the plot to see which the object 
matches most closely. Figure~\ref{fig:Sixspectra} and Figure~\ref{fig:spec2} show the closest-matching template spectra for each object. A type is derived from the index values by taking their median. This is then averaged with the template spectrum type; the advantage of this process is that it accounts for both broad spectral structure and also for index-based measurements.

Only the indices in the $J$- and $H$-bands have been used for the purpose of typing the T dwarfs in this study, as unusual $K$-band spectroscopic morphology is one of the defining characteristics
by which they were selected, potentially invalidating the use of this spectral region as a means of obtaining spectral types consistent with those for normal T dwarfs. In addition, we do not have $K$-band spectra for all objects.
 Therefore it was felt acceptable to neglect the $K$-bands for spectral-typing purposes. However the index values are shown in Table~\ref{tab:indices} for reference. This table also
summarises the bands used in our typing and the template spectra that were selected. Spectra for all candidates are plotted in figures Figure~\ref{fig:Sixspectra} and Figure~\ref{fig:spec2}.

In cases where spectroscopy is available over the full $JHK$ region, the uncertainties in our types are $\pm0.5$ subtypes. However, for some objects we only have $J$-band data or $JH$-data.
 Also, our comparison template spectra allow us to type an object to $\pm1$ subtype accuracy, not $\pm0.5$. Therefore we have rounded our accuracies to $\pm1$ subtype, although index types themselves are
 generally more accurate. 

\begin{landscape}
 \begin{table}
\begin{tabular}{lllllllll} 
\hline
Object & $H_{2}O-J$ & $CH_{4}-J$ & $H_{2}O-H$ & $CH_{4}-H$ & $CH_{4}-K$ & Median & Template & Adopted \\
       &            &            &            &            &            & type   & spectrum & type     \\
\hline
ULAS~J0842+0936 & 0.17$\pm0.02$ (T5/T6)   & 0.35$\pm0.02$ (T5/T6)   &                          &                         &                      & T5.5 & T6  & T6$\pm1$ \\
ULAS~J0926+0835 & 0.32$\pm0.02$ (T4/T5)   & 0.53$\pm0.02$ (T3/T4)   &                          &                         &                      & T4   & T4  & T4$\pm1$ \\
ULAS~J0958-0039 & 0.16 (T6)               & 0.34 (T6)               & 0.39 (T4)                & 0.41 (T5)               & 0.14 (T6)            & T6   & T6  & T6p$\pm1$ \\
ULAS~J1012+1021 & 0.202$\pm0.008$ (T5)    & 0.366$\pm0.009$ (T5)    &                          &                         &                      & T5   & T6  & T5.5$\pm1$ \\
ULAS~J1018+0725 & 0.36 (T4)               & 0.44 (T5)               & 0.38 (T4)                & 0.45 (T5)               & 0.15 (T6)            & T5   & T5  & T5$\pm1$ \\
ULAS~J1233+1219 & 0.365$\pm0.007$ (T4)    & 0.528$\pm0.007$ (T3)    &                          &                         &                      & T3.5 & T4  & T4$\pm1$ \\
ULAS~J1303+0016 & 0.11 (T7)               & 0.32 (T6)               & 0.34 (T5)                & 0.37 (T5)               & 0.03 (T7)            & T6   & T6  & T6p$\pm1$ \\
ULAS~J1319+1209 & 0.233$\pm0.009$ (T5)    & 0.322$\pm0.006$ (T6)    & 0.45$\pm0.02$ (T3/T4)    &                         &                      & T5   & T5  & T5$\pm1$ \\
ULAS~J1320+1029 & 0.247$\pm0.003$ (T5)    & 0.479$\pm0.004$ (T4)    & 0.318$\pm0.004$ (T6)     &                         &                      & T5   & T5  & T5$\pm1$ \\
ULAS~J1501+0822 & 0.29 (T5)               & 0.43 (T5)               & 0.41 (T5)                &                         &                      & T5   & T4  & T4.5$\pm1$ \\
Hip~73786B      & 0.127$\pm0.004$ (T6/T7) & 0.383$\pm0.003$ (T5)    & 0.307$\pm0.006$ (T6)     & 0.355$\pm0.009$ (T5/T6) & 0.45$\pm0.03$ (T3)   & T5.5 & T6  & T6p$\pm1$ \\
ULAS~J2320+1448 & 0.206$\pm0.004$ (T5)    & 0.396$\pm0.004$ (T5)    & 0.303$\pm0.006$ (T6)     & 0.292$\pm0.003$ (T6)    & 0.172$\pm0.004$ (T6) & T6   & T6  & T6$\pm1$ \\
\hline
\end{tabular}
\caption{Summary of the objects' spectral type indices, as per \citet{Burg_2006}. The spectral types assigned from those indices are shown in parentheses after the number.
 Where the error on an index is consistent with more than one type, this is shown with a slash, for instance T4/T5 would indicate an object whose relevant index could be consistent with
 either value. Uncertainties in typing are assigned based on the types from different indices. The Template Spectrum column refers to the type of the spectral standard that best matches the shape
 of the spectrum. The Adopted Type column refers to the spectral type that was derived for each object. Errors are shown where error spectra are available. Error spectra are not available for
 ULAS~J0958-0039, ULAS~J1018+0725, ULAS~J1303+0016 or ULAS~J1501+0822.
Final spectral types are arrived at by averaging the median of the $J$, $H$-band indices with the type from the template comparison. This is done to give equal weighting to broad structure
 and numerical indices.
In all cases, the $H_{2}O-J$ and $CH_{4}-J$ indices were used. Where $H$ and $K$-band spectra were available, $H_{2}O-H$, $CH_{4}-H$ and $CH_{4}-K$ indices were also used.
\label{tab:indices}}
 \end{table}
\end{landscape}

\subsection{Notes on unusual objects}
Some objects show a large scatter in their indices, sometimes varying by as much as three subtypes between indices. This can be seen in Table~\ref{tab:indices},
 where the type inferred from each index is shown in parentheses after the value.

Objects with $H_{2}O-H$-early peculiarity \citep{Burningham_2010} have a $H_{2}O-H$ index at least 2 subtypes earlier than the $H_{2}O-J$ index. Amongst our objects, ULAS~J0958-0039 displays this behaviour, with an assigned type of T6
 and a $H_{2}O-H$ type of T4 and a $H_{2}O-J$ type of T6. ULAS~J1303+0016 also shows this behaviour. There is an additional borderline case, ULAS~J1319+1209, which has a $H_{2}O-J$ type of T5
 and a $H_{2}O-H$ type of T3/T4.

It is possible that the cause for these peculiarities may lie in the physics of low-metallicity or high surface gravity atmospheres. However we do not see this behaviour consistently across our sample, and anyway without derived metallicities or surface gravities for our objects it is not currently possible to evaluate this idea.
 Binarity has been ruled out as possible cause, however \citep{Burningham_2010}.

Lastly, ULAS~J1018+0725 shows some differences both from the standards and from the rest of the sample. Its $Y$-band peak is narrower relative to the standards and there is less flux bluewards of 1.1$\mu$m. Also, its $K$-band peak actually appears enhanced relative to the standards.
 These may in fact be evidence of high metallicity and/or low gravity, as suggested in \citet{Lodieu_2007}. Its $H-K = -0.3\pm0.2$, which is not particularly blue relative to the rest of the sample. Given the error on its colours, it may be a higher-metallicity object that has scattered into our colour-selection area.

\section{Distance estimates}
\label{sec:Dist}
The distances of the candidates were estimated using the relationships between spectral types and absolute magnitudes from \citet{Marocco_2010}. The $J$-band was used for distance estimates,
 as it is considered least affected by gravity or metallicity. The results are summarised in Table~\ref{tab:Kinematics}. The relationship used
 is not based on metal-poor T dwarfs. Metallicity effects may change the absolute magnitudes of T dwarfs in the $J$-band.
 However, we continue to use the \citet{Marocco_2010} relationships, making the working assumption that they are applicable. Errors on the distances were estimated using the uncertainties in spectral type, the uncertainties in $J$-band magnitude and the scatter in the relation.

 \begin{table*}
\small{
\begin{tabular}{lllllllll}       
\hline
 Name & UKIDSS date & Epoch diff  & Follow-up     & n$_{ref}$ & T dwarf SNR & X,Y RMS   & $\mu_{\alpha cos \delta}$, $\mu_{\delta}$ & S1, S2 \\
      &             & (yr)        &    source     &           &             & (pixels)  & (mas\,yr$^{-1}$)                          & (pixels) \\
\hline
ULAS~J0842+0936 & 2007/02/16 & 0.90603 & UFTI       & 11 & 9.0, 33.3  & 0.111, 0.306 & -176$\pm209$,  +36$\pm218$  &  5.1, 8.3 \\
ULAS~J0926+0835 & 2007/01/22 & 1.02155 & EMMI       & 17 & 6.6, 13.7  & 0.107, 0.156 & -472$\pm144$, -438$\pm146$  &  7.7, 8.4 \\
ULAS~J0958-0016 & 2005/12/30 & 2.08705 & EMMI       & 15 & 7.2, 17.9  & 0.148, 0.174 & +43$\pm99$, +31$\pm100$     &  4.1, 5.0 \\
ULAS~J1012+1021 & 2007/12/02 & 0.61579 & WFCAM      & 23 & 32.3, 43.5 & 0.085, 0.093 & -234$\pm87$, -631$\pm87$    &  4.4, 5.3 \\
ULAS~J1018+0725 & 2005/12/28 & 1.05775 & UKIDSS     & 26 & 18.2, 18.5 & 0.104, 0.120 & +181$\pm48$, +60$\pm50$     &  3.2, 3.2 \\
ULAS~J1233+1219 & 2007/03/05 & 0.74209 & EMMI       & 25 & 11.9, 22.2 & 0.105, 0.134 & +198$\pm189$, +168$\pm190$  &  3.0, 6.7 \\
ULAS~J1303+0016 & 2005/06/12 & 4.01232 & UKIDSS     & 23 & 4.9, 3.5   & 0.130, 0.106 & -18$\pm138$, -220$\pm137$   &  3.1, 4.2 \\
ULAS~J1319+1209 & 2007/04/16 & 0.79740 & EMMI       & 23 & 8.5, 12.9  & 0.089, 0.139 & -525$\pm72$, +111$\pm75$    &  5.2, 8.0 \\
ULAS~J1320+1029 & 2007/02/25 & 1.17975 & EFOSC2     & 21 & 16.1, 8.3  & 0.151, 0.151 & +9$\pm242$, -60$\pm243$     &  5.2, 11.8 \\
ULAS~J1501+0822 & 2006/07/23 & 0.61084 & LIRIS      & 12 & 8.0, 12.5  & 0.310, 0.237 & -136$\pm180$, -113$\pm249$  &  8.5, 4.2 \\
Hip~73786B      & 2008/04/17 & 1.23460 & WFCAM      & 45 & 27.8, 20.4 & 0.095, 0.093 & -661$\pm60$, -510$\pm59$    &  4.8, 7.5 \\
ULAS~J2320+1448 & 2008/06/27 & 7.83025 & SDSS       & 14 & 18.9, 11.3 & 0.093, 0.104 & +399$\pm26$, +122$\pm26$    &  4.7, 2.2 \\
\hline
\end{tabular}
}
\caption{Summary of the calculations for object proper motions.
UKIDSS date is the date of the $J$-band LAS image.
 n$_{ref}$ refers to the number of reference stars used for each object.
 The SNR for the T dwarf from UKIDSS is shown first, followed by the SNR from the second epoch. 
X, Y RMS refers to the RMS scatter about the co-ordinate transforms.
 The next two columns show the calculated proper motions for RA
 and declination, in mas\,yr$^{-1}$.  S1 is the seeing in the LAS image, S2 the seeing in the follow-up image. 
Coordinates are transformed from the second-epoch image to the UKIDSS $J$-band image for each object. The UKIDSS $J$-band pixel scale is 0.2 arcseconds per pixel.
\label{tab:pmstuff}}
 \end{table*}

\section{Proper motions}
\label{sec:PM}
 The photometric follow-up observations that were carried out provided a second epoch of imaging data for each object.
 We used Image Reduction and Analysis Facility \citep[IRAF; see][]{Tody_1986} task {\sc GEOMAP} to derive spatial transformations into
the UKIDSS
 LAS $J$-band image for each object, based on the positions of reference stars identified in the follow-up and UKIDSS images. We used the IRAF package {\sc GEOMAP} for the co-ordinate transformation.
The fitting geometry used was general, with a polynomial function of order 3 in x and y.
In every case the follow-up image was used as the reference image for the co-ordinate transform. 
Followup images were obtained from a variety of instruments, including the UKIRT Fast-Track Imager \citep[UFTI; see][]{Roche_2002}, the Long-slit Intermediate Resolution Infrared Spectrograph \citep[LIRIS; see][]{Manchado_2003}, the Wide Field Infrared Camera For UKIRT \citep[WFCAM; see][]{Casali_2007}, the ESO Multi-Mode Instrument \citep[EMMI; see][]{Dekker_1986} and the ESO Faint Object Spectrograph and Camera \citep[EFOSC2; see][]{Buzzoni_1984}.

We then transformed the
pixel coordinates of the targets in the follow-up
   images into the LAS image, using {\sc GEOXYTRAN}, and calculated their change in position
(relative to the reference stars) between the two epochs.

The uncertainties associated with our proper motion measurement primarily come from the spatial transformations,
 and the accuracy with which we have been able to measure the position of the targets (by centroiding) in the image data.
 Then the instrument pixel scale was used in each case to convert the results into arcseconds. The UKIDSS $J$-band pixel scale is 0.2 arcseconds per pixel.
 The dates of the observations were then used to calculate a proper motion. These results are summarised in Table~\ref{tab:pmstuff}.

Errors on proper motions were calculated by summing in quadrature the centroiding uncertainty and the RMS about the fit of the co-ordinate transform and the epoch difference.
The uncertainty in the centroiding was estimated by obtaining the scatter of the centroids for a large number of simulated stellar images constructed using the observed signal-to-noise and seeing for each target observation.
Centroiding uncertainties were calculated through
simulated data with a Guassian PSF and with appropriate Poisson noise added. The
availability and quality of the measured proper motions of
the T dwarfs is affected by various factors including the
time between epochs, the number of usable reference stars in the images and the S/N
of both the T dwarf and the reference stars. The centroiding uncertainties (in pixels) varied between a minimum of 0.04 pixels for ULAS~J1233+1219 and a maxmum of 2.38 pixels for ULAS~J1320+1029. The mean value was 0.54 pixels.

ULAS~J2320+1448 was measured with the method of \citet{Zhang_2009}, with its UKIDSS and SDSS images \citep[SDSS; see][]{York_2000}. Due to the much longer baseline, this has resulted in a lower error for this object's proper motion compared to the
 rest of the sample.

 \begin{table*}
\begin{tabular}{lccccccc}
\hline
Name             & Distance   & V$_{tan}$       & $U$            & $V$            & $l$    & $b$   & $[U^{2}+(V+35)^{2}]^{1/2}$ \\
                 & (pc)       & (km\,s$^{-1}$)  & (km\,s$^{-1}$) & (km\,s$^{-1}$) & (deg)  & (deg) & (km\,s$^{-1}$) \\
\hline
ULAS~J0842+0936  &  50$\pm9$  & 43$\pm40$       & 38$\pm36$      & -1$\pm44$       & 216.89  & 28.94  & 51 \\
ULAS~J0926+0835  &  70$\pm11$ & 213$\pm59$      & 62$\pm40$      & -140$\pm46$     & 223.84  & 38.16  & 122 \\
ULAS~J0958-0039  &  65$\pm9$  & 16$\pm22$       & 3$\pm27$       & -3$\pm27$       & 239.49  & 40.02  & 32 \\
ULAS~J1012+1021  &  28$\pm5$  & 89$\pm20$       & -1$\pm11$      & -87$\pm17$      & 229.37  & 49.09  & 52 \\
ULAS~J1018+0725  &  44$\pm7$  & 39$\pm12$       & -16$\pm10$     & 5$\pm9$         & 234.27  & 48.75  & 43 \\
ULAS~J1233+1219  &  50$\pm8$  & 62$\pm40$       & -11$\pm45$     & 45$\pm45$       & 286.16  & 74.61  & 81 \\
ULAS~J1303+0016  &  67$\pm12$ & 70$\pm42$       & -20$\pm44$     & -68$\pm45$      & 309.34  & 63.00  & 39 \\
ULAS~J1319+1209  &  75$\pm12$ & 192$\pm40$      & 183$\pm37$     & -92$\pm29$      & 328.20  & 73.62  & 192 \\
ULAS~J1320+1029  &  46$\pm7$  & 13$\pm37$       & 0$\pm51$       & -21$\pm52$      & 326.89  & 71.98  & 19 \\
ULAS~J1501+0822  &  60$\pm10$ & 51$\pm49$       & 12$\pm52$      & -61$\pm63$      & 7.70    & 53.97  & 41 \\
Hip~73786B       &  22$\pm4$  & 87$\pm25$       & -48$\pm7$      & -75$\pm4$       & 4.90    & 51.66  & 62 \\
ULAS~J2320+1448  &  24$\pm5$  & 48$\pm10$       & 55$\pm10$      & -20$\pm3$       & 92.45   & -42.62 & 57 \\
\hline
\end{tabular}
\caption{This table shows object distances in parsecs, estimated by spectral type, using the relationships between $J$-band type and absolute magnitude
 from \citet{Marocco_2010}. The distance range shows the distances for the earliest and latest spectral types
 that are consistent with each object's spectral type range. It also shows the estimated tangential velocities, and their errors. With the exception of HIP~73786B, all objects are assumed to have $v_{rad}=0$ km\,s$^{-1}$.
\label{tab:Kinematics}}
\end{table*}

\section{Kinematics}
\label{sec:Kinematics}

We have estimated tangential velocities for our targets using the proper motion and distance estimates, listed in Table~\ref{tab:Kinematics}. Errors in tangential velocity and $U$ and $V$ were estimated using a Monte Carlo script, accounting for errors in proper motion and in the distance estimates.
 The Monte Carlo used 10000 loop iterations per object.

Two objects were found to have $V_{tan}$ greater than 100 km\,s$^{-1}$, suggestive of halo kinematics: ULAS~J0926+0835 with $V_{tan} = 213\pm59$ km\,s$^{-1}$ and ULAS~J1319+1209 with $V_{tan} = 192\pm40$ km\,s$^{-1}$.

\begin{figure*}
\includegraphics[height=500pt, angle=90]{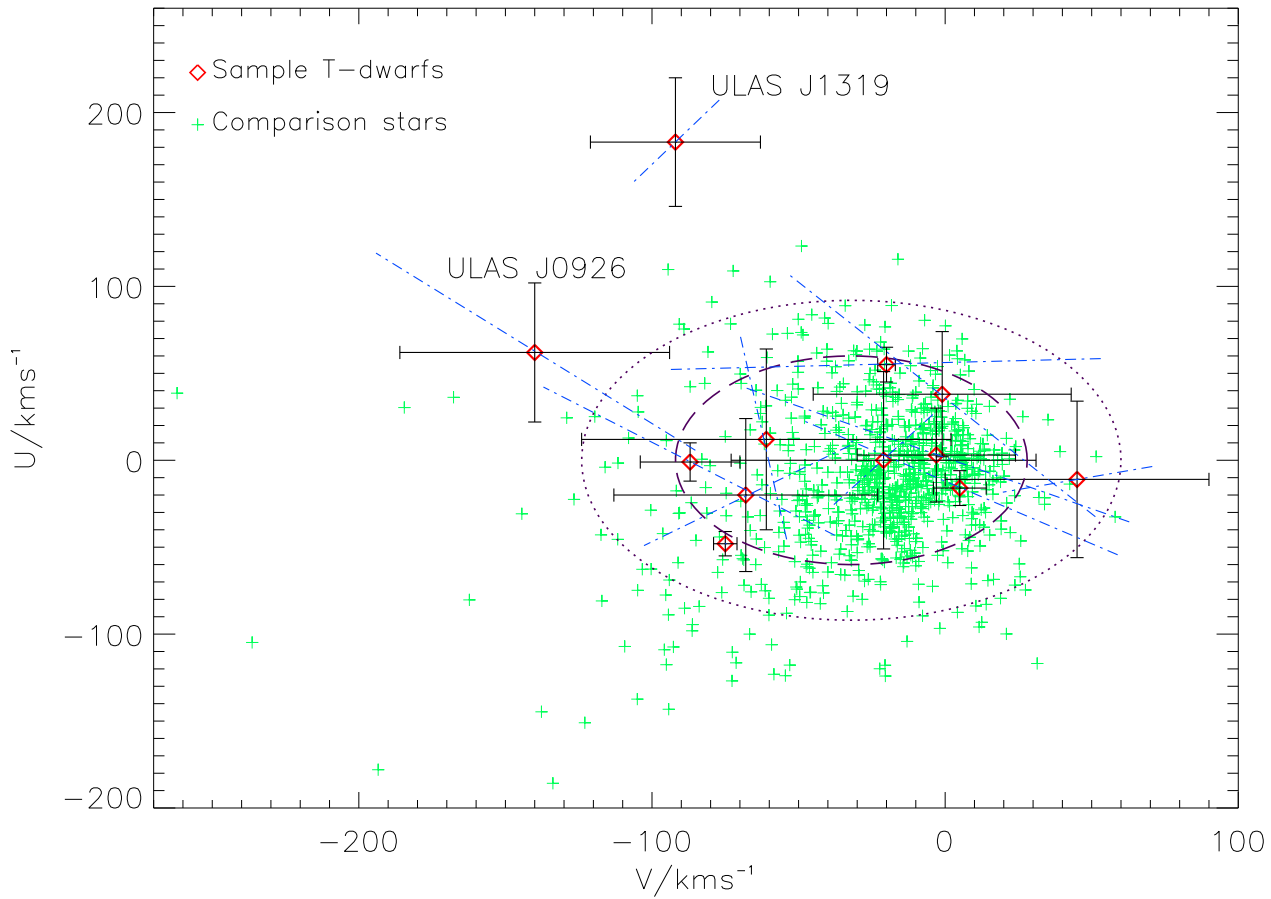}
\caption{$U$ and $V$ components plotted for our sample (red diamonds) and a stellar background (green crosses). The stellar background is from \citet{Soubiran_2008}. Error bars are based on distance errors and proper
 motion errors. The purple dashed line marks the 1$\sigma$ velocity ellipsoid for disc stars and the dotted line marks the 2$\sigma$ disc ellipsoid \citep[see ][]{Chiba_2000,Reid_2001}. The blue angled, dot-dashed lines represent $U$ and $V$ values in the cases that $V_{rad}=\pm100$ km\,s$^{-1}$ respectively.}
\label{fig:UVplot}
\end{figure*}

As the LAS is oriented out of the galactic plane, some insight can be gained into our targets' $U$ and $V$ space velocity components. For targets at high galactic latitudes, the radial velocity will consist mostly of $W$ motion, with small $U$ and $V$ components. 
Therefore, in such cases it is possible to assess $U$ and $V$ to some extent without having a radial velocity. So, for all objects, proper motions, distance
 and co-ordinates were transformed into $UVW$ components. This was done assuming $V_{rad}=0$ for all objects except Hip~73786B, where the values for Hip~73786A were used. The results are summarised in Table~\ref{tab:Kinematics}.
 The results are also displayed in Figure~\ref{fig:UVplot}. The scheme we have used is a left-handed one, such that $U$ is positive toward the galactic anti-centre and $V$ positive in the direction of galactic rotation.
To account for the possible effects of radial velocity, we have also calculated $U$ and $V$ for each object for the possibilities of $V_{rad}=100$ and $-100$ km\,s$^{-1}$. This is shown in Figure~\ref{fig:UVplot} by the blue dot-dashed lines running through each object. 

Four objects (ULAS~J1233+1219, ULAS~J1303+0016, ULAS~J1319+1209 and ULAS~J1320+1029) have $b > 60$ degrees, and their $U$ and $V$ values are the best constrained in our study. Of these, ULAS~J1233+1219 displays kinematics most likely to be young disc values, although its $U$,$V$ error bars do extend beyond the outer disc ellipsoid. Unless its $V_{rad}$ is outside the $\pm 100$km\,s$^{-1}$ indicated in Figure~\ref{fig:UVplot}, these kinematics seem to argue against it being a halo object. If, it is not a halo object, then its blue near-infrared colours come as a surprise; ULAS~J1233+1219 has the bluest $J-K$ in the entire sample ($J-K = -1.2 \pm0.1$). This suggests a strongly-depressed $K$-band and is notable
 in light of its apparently disc-like tangential velocity (although no $K$-band spectrum has yet been obtained for this object). This situation is reminiscent of the only T~dwarf known to be bluer in $J-K$, SDSS~J1416+1348B \citep[see ][]{Ben_2010,Scholz_2010a}, with $J-K = -1.67$. THe SDSS~J1416+13AB system has velocities consistent with the young disc.

 \citet{Reid_2001} set the criterion that objects that satisfy

\begin{equation}
[U^{2}+(V+35)^{2}]^{1/2} > 94 
\end{equation} km\,s$^{-1}$

are outside of the $2 \sigma$ velocity region encompassing young and old-disc stars as described by \citet{Chiba_2000} and \citet{Reid_2001}. ULAS~J0926+0835 and ULAS~J1319+1209 satisfy this condition (see Table~\ref{tab:Kinematics}), further highlighting their likely halo membership. 

Two objects stand out on Figure~\ref{fig:UVplot} as halo candidates. ULAS~J1319+1209 has $U$ and $V$ ($U=183\pm37$, $V=-92\pm29$ km\,s$^{-1}$) that place it somewhat outside the 2$\sigma$ disc ellipsoid. Also, due to its high galactic latitude, there is little contribution to these from radial velocity. ULAS~J0926+0835 has a more poorly-constrained $U$ and $V$, but may also be halo-like, with $U=62\pm40$, $V=-140\pm46$ km\,s$^{-1}$. The $\pm100$ km\,s$^{-1}$ line for ULAS~J0926+0835 intersects the disk rings, unlike ULAS~J1319+1209. This means that ULAS~J1319+1209 is the stronger halo candidate. The weaknesses associated with ULAS~J0926+0835's candidature also highlights the need for radial velocities for these objects.

Of the remaining objects, the $U$ and $V$ components seem most likely to be within the young-disc range, including ULAS~J1018+0725 (which supports its interpretation as a metal-rich object that has scattered into the sample). However, with the exception of Hip~73786B, the possibility of large radial velocity components for these objects prevents us from ruling them out as halo candidates.

It is interesting to note that our strongest halo candidates do not appear to be the bluest objects in the sample. In $J-K$, six other objects show a bluer colour but have more modest $V_{tan}$ and apparently disc-like $U$ and $V$. At any given spectral type, the combined effect of greater age (and thus higher gravity) and lower metallicity should result in bluer $J-K$ colours for substellar halo members than for the members of the younger disc population. Surprisingly, for both of our kinematic halo candidates several objects of similar spectral type are seen to be 2$\sigma$ bluer in $J-K$. However, better photometry would be required to confirm this trend for our halo candidates.

\section{Hip 73786A and B}
\label{LHS3020}

 \begin{table}
\begin{tabular}{ll}
\hline
 Property & Value \\
\hline
 R.A. (J2000.0)       & 15:04:53.5267 \\
 Dec. (J2000.0)       & +05:38:17.150 \\
 Distance (pc)        & 19.2$\pm0.7^{a}$ \\
 Spectral type        & K5$^{b}$ \\
 T$_{eff}$/K          & 4450$\pm50^{b}$ \\
 log $g$              & 4.80$\pm0.1^{b}$ \\
 $[Fe/H]$             & -0.30$\pm0.1^{b}$ \\
 $B$/mag              & 11.16$\pm0.02^{c}$ \\
 $V$/mag              & 9.85$^{d}$ \\
 $R$/mag              & 9.0$\pm0.3^{e}$ \\
 $I$/mag              & 8.2$\pm0.3^{e}$ \\
 $J_{2MASS}$/mag      & 7.26$\pm0.02^{f}$ \\
 $H_{2MASS}$/mag      & 6.66$\pm0.02^{f}$ \\
 $K_{2MASS}$/mag      & 6.47$\pm0.02^{f}$ \\
 $\mu_{\alpha cos \delta}$/mas\,yr$^{-1}$ & -606.8 $\pm1.8^{a}$ \\
 $\mu_{\delta}$/mas\,yr$^{-1}$ & -507.6 $\pm1.8^{a}$ \\
 $V_{rad}$/km\,s$^{-1}$ & -68$\pm10^{g}$ \\
 $U$/km\,s$^{-1}$       & -48$\pm7$ \\
 $V$/km\,s$^{-1}$       & -75$\pm7$ \\
 $W$/km\,s$^{-1}$       & -44$\pm8$ \\
\hline
\multicolumn{2}{l}{$^{a}$: \citet{Perryman_1997} } \\
\multicolumn{2}{l}{$^{b}$: \citet{Cenarro_2007} } \\
\multicolumn{2}{l}{$^{c}$: \citet{Weis_1993} } \\
\multicolumn{2}{l}{$^{d}$: \citet{Evans_1967} } \\
\multicolumn{2}{l}{$^{e}$: \citet{Monet_2003} } \\
\multicolumn{2}{l}{$^{f}$: \citet{Cutri_2003} - 2MASS point-source catalogue} \\
\multicolumn{2}{l}{$^{g}$: \citet{Barbier_2000} } \\
\hline
\end{tabular}
\caption{Properties of Hip 73786A.
\label{tab:LHS3020A}}
 \end{table}

\begin{figure*}
\includegraphics[height=500pt, angle=90]{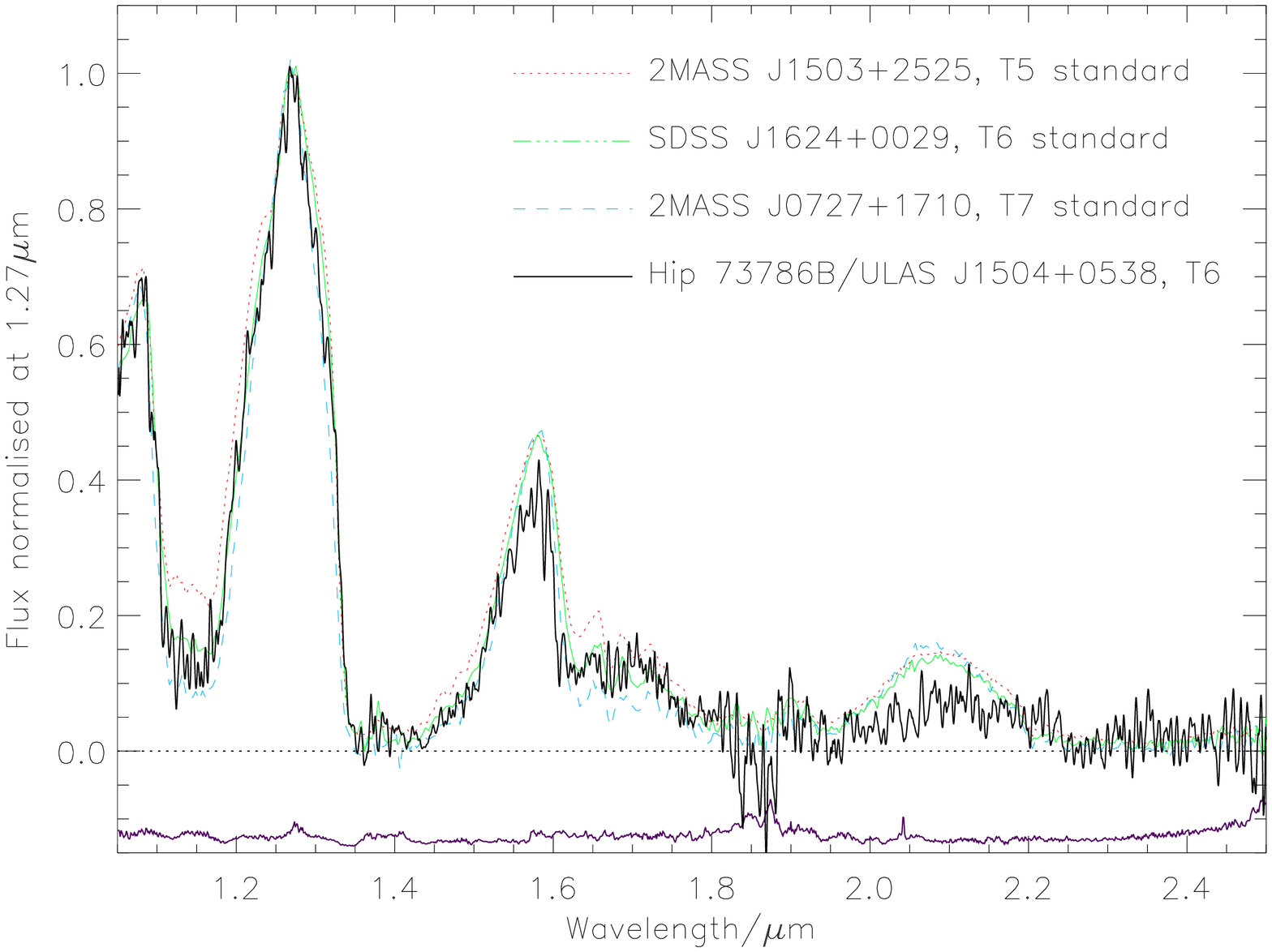}
\caption{Spectrum for Hip 73786B, plotted along with T7, T6 and T5 standard spectra. The T7 and T6 spectra are from \citet{Burg_2006} and the T5 is from \citet{Burg_2004}. The thicker black line is the object spectrum. All spectra are normalised to 1.27$\mu$m and have been placed onto the object's wavelength scale.
 It can be seen that Hip 73786B's $K$-band is deeply depressed
relative to the standards. Also, its $H$-band shows evidence for similar depression. The $J$-band shape is broadly similar to the standards, however.
 The error spectrum, the purple line at the bottom, has been offset by -0.15 for clarity.}
\label{fig:J1504_spec}
\end{figure*}

\citet{Scholz_2010} identified Hip~73786B as a common proper motion companion to Hip~73786, a K5 dwarf at a distance of 19.2pc \citep{Perryman_1997}. Our search for blue T~dwarfs and kinematic analysis also independently identified this pair, and for completeness we describe our analysis below.

 We find $\mu_{\alpha cos \delta}=-627\pm160$ mas\,yr$^{-1}$ and 
$\mu_{\delta}=-457\pm59$ mas\,yr$^{-1}$ for Hip~73786B. These values are consistent with both \citet{Scholz_2010} and with Hip 73786. Thus we concur that the two objects share common motion.

We consider the possibility that the two components of Hip~73786B and Hip~73786A are simply randomly aligned on the sky with the same proper motions. To control for this, we use the method of \citet{DayJones_2008}.
 This was begun by taking objects with proper motions ($>$10 mas\,yr$^{-1}$) from SuperCOSMOS in a square degree of sky surrounding Hip~73786B. In this region of sky 29,249 objects had proper motions. We then placed these onto a colour-magnitude diagram ($B$ vs $B-R$) and selected only those with a consistent magnitude to be at a common distance with Hip~73786B.
Thirty-nine objects had colours and magnitudes consistent with being at the same distance as our T dwarf. We then placed these objects on a vector point diagram and compared their proper motion to that of Hip~73786B. None of the thirty-nine objects had proper motions consistent to within 2$\sigma$ of our T dwarf, suggesting that the likely contamination of objects with the same distance and proper motion to Hip~73786B is $<$1/39.

We next calculated the associated volume of space in which we could observe contaminant objects. Allowing for the possibility of unresolved binarity we can estimate a liberal range of plausible distances to the T dwarf as 18-31~pc. Considering the separation of 63.8\arcsec, we estimate that the pair of objects share a volume of $<0.8$pc$^{3}$.
As Hip 73786 is a K5 star, we consider the number density of K type stars,
 which is 0.0384 pc$^{-3}$ \citep{McCuskey_1983}. Therefore we would expect to observe $0.0384 \times 0.8=0.031$ K-type stars within this volume. Overall then, the chance of a spurious common proper motion main sequence companion to Hip~73786B is $<7.9 \times$ 10$^{-4}$.
 Indeed, we would expect $<0.15$ such spurious companions amongst the entire, $\sim$200 strong, sample of published T dwarfs. We thus consider that Hip~73786B and Hip 73786A are a genuinely associated system of common origin, confirming the analysis of \citet{Scholz_2010}

The two objects' apparent separation is $63.8"$. At a distance of 19.2 pc this corresponds to a projected separation of $\sim1230$ AU. Considering also the probabilities discussed above, we assume that the two objects share a common origin, being either a binary or a co-moving pair.
 We also posit the working assumption that common formation history implies a common composition between Hip~73786A and Hip~73786B, allowing for tighter constraints on the T dwarf's
 physical characteristics.

Hip 73786 is known to have a metallicity of $[Fe/H]=-0.3\pm0.1$ \citep{Cenarro_2007}, clearly identifying it as a metal-poor star. Although this metallicity remains within the young-disc range of $[Fe/H]=-0.4$ to $+0.3$ \citep{Nissen_1999}, nonetheless this 
 would suggest that Hip~73786B represents a new metal-poor T dwarf benchmark.

As Hip 73786A has a radial velocity $v_{rad}$ = -68$\pm10$ km\,s$^{-1}$ \citep{Barbier_2000}, $UVW$ space motion components were calculated. We find that $U=-47.9\pm6.6$ km\,s$^{-1}$,
 $V=-74.7\pm6.5$ and $W=-43.9\pm8.4$.

Hip 73786A shows some evidence of chromospheric activity. \citet{Gray_2003} list a value of $log\,R'_{hk}=-4.76$ for the primary. \citet{Mamajek_2008} gives a relationship between $log\,R'_{hk}$ and stellar age. Using this
 and the value from \citet{Gray_2003}, we find an age of 2.0 Gyr for Hip 73786A. However, several factors can affect $log\,R'_{hk}$. \citet{Henry_1996} list binarity as one such factor, and there is a suggestion in \citet{Luyten_1979} that Hip 73786A may itself be a close binary, although no other references exist to support this. 
Also, with regard to our Sun, the value
 of $log\,R'_{hk}$ varies depending on the point in the solar cycle at which it is measured. The Sun can vary between an estimated extreme of -5.10 during solar minimum to a peak of -4.75 during solar 
maximum \citep{Henry_1996}. These fluctuations correspond to an age in the range of 2.2 to 8.0 Gyr, as opposed to the actual value of 5 Gyr. With only one available measurement, it is unknown whether or not Hip~73786A presents a 
similar activity cycle. If it does, then the example of the Sun shows that the age of the Hip~73786 system may be different from the estimated value.

Another possible source of uncertainty here is that \citet{Jenkins_2008} find an offset of $\sim$0.1~dex between their analysis of typical FGK dwarfs and those of Gray et al. Jenkins et al. also show 
this offset is present in cross-matched samples of FGK stars between Gray et al. and Henry et al.  Since this appears to be a systematic offset we can apply this correction to the value of Hip 73786A 
found by Gray et al.  Using this correction would suggest a $log\,R'_{hk}=-4.66$ for Hip 73786A, which in turn feeds through to an age estimate of $\sim1.6$ Gyr, slighter lower than the adopted 
age if we accept the Gray et al. value at face value. 

\citet{Holmberg_2009} find a mean $[Fe/H]=-0.24$ for stars older than 4.0 Gyr. The age-velocity relation they find would also suggest an age $>$8 Gyr for the Hip 73786 system. These factors may suggest an older age than the one we derive through chromospheric activity.
However, both of these relations are subject to a large
 amount of scatter and kinematic relations are best applied to populations, not individual objects. The metallicity could also be consistent with an age below 4.0 Gyr, which limits the usefulness of this property for age detemination Therefore we suggest that the Hip 73786 system has a minimum age constraint of 1.6 Gyr, but we are currently unable to estimate an upper limit.

As a control for the possibility of binarity, high-resolution imaging was undertaken for Hip 73786A.
Observations were carried out with FastCam, the 'Lucky Imaging' facility
\citep{Oscoz_2008} installed as a common-user instrument on the Carlos S\'anchez
Telescope in the Teide Observatory, Tenerife, Canary Islands. FastCam is
equipped with a low read-out noise L3CCD Andor 512$\times$512 camera
with a pixel scale of 42.2 milli-arcsec (mas), yielding a field-of-view
of approximately 21.6 by 21.6 arcsec.

HIP\,73786 was observed on 24 July 2010 in the $I$-band filter under a
natural seeing of 1.0 arcsec, clear conditions, and full moon. The total
exposure time was divided up into 1000 images of 50ms repeated five times.
The achieved resolution is of the order of 0.2 arcsec, corresponding to a
projected physical separation of less than 4 AU, assuming a distance of 19 pc
for the target \citep{Perryman_1997}. The data reduction of the raw images was done with the automatic pipeline
distributed by the FastCam team and developed by the Universidad Polyt\'ecnica
de Cartagena\footnote{Details on FastCam at http //www.iac.es/proyecto/fastcam/}.
This reduction involves bias and flat-field correction and the best
15\% exposures were selected to achieve a near to diffraction-limited image.  
From this analysis we find no companion at the observed resolution
around HIP 73786 down to 4-5 mag at a separation of 1 arcsec from the target.

%
%
\begin{figure}
 \centering
 \includegraphics[height=250pt, angle=0]{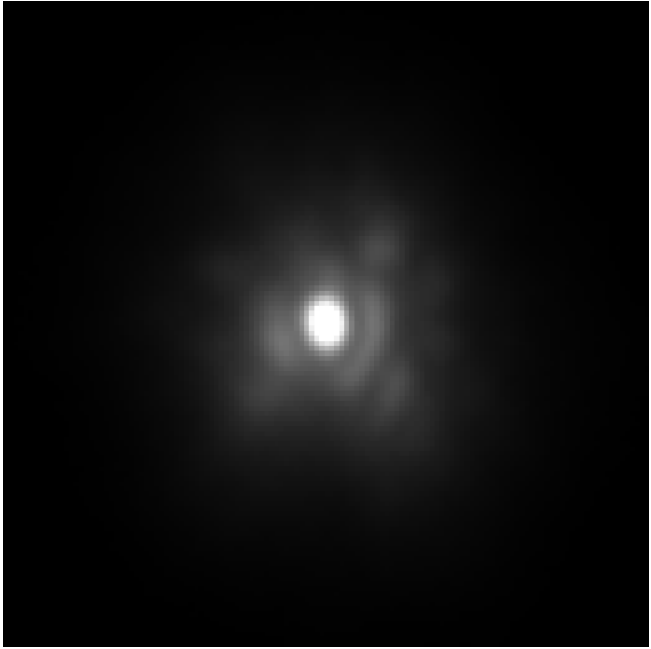}
 \caption{Image of HIP 73786 taken in the $I$-band filter with
FastCam installed on the Carlos S\'anchez Telescope in the Teide
Observatory. The pixel scale 42.2 mas and the field-of-view is
2.7 by 2.7 arcsec aside. East is left and North is up. No companion
is detected around HIP 73786 up to 10 arcsec from the primary using
the 15\% and 50\% of the images.}
\label{fig:LuckyImage}
\end{figure}
%


\section{Summary}
\label{sec:Summary}
We present data for twelve UKIDSS T dwarfs, including 11 from DR4 and one object from DR5. We use follow-up data to compute proper motions and
 tangential velocities, and find a range of values, ULAS~J0926+0835 and ULAS~J1319+1209 with $V_{tan} > 100$ km\,s$^{-1}$, which may be indicative of halo kinematics. Confirmation or definitive rejection
of these objects will need measurements of $V_{rad}$ and calculation of full $UVW$ space motions.

Spectra are presented for all candidates. Spectral typing is undertaken on all of them using the scheme of \citet{Burg_2006}. A $\pm2$ subtype variance is noted on the indices of some of our objects.

Hip~73786B from UKIDSS DR5 is confirmed to be associated with the K5 dwarf Hip 73786A, and we find it to be a T6p dwarf. Hip~73786B is found to be at least $1.6$ Gyr old, constrained by chromospheric activity in the primary. As Hip 73786A is known to be of low metallicity, this system represents and important addition to the sample of T~dwarf benchmark systems. 
 We note that one object (ULAS~J1018+0725) appears to be a metal-rich or low-gravity, disc dwarf that has scattered into our sample. Of the rest, we find two objects with strong kinematic evidence of halo membership (ULAS~J0926+0835 
and ULAS~J1319+1209). Most interestingly we note that our bluest objects are not the best halo candidates, although the current velocity and colour uncertainties means this observation is only suggestive and must be verified with more precise proper motion and radial velocity measurements.

\section{Acknowledgements}
This work has benefited from data from the UKIRT Infrared Deep Sky Survey (http://surveys.roe.ac.uk/wsa/pre/index.html).

This research has benefited from the SpeX Prism Spectral Libraries, maintained by Adam Burgasser at http://www.browndwarfs.org/spexprism

This research has made use of the SIMBAD database, operated at CDS, Strasbourg, France.

Funding for the SDSS and SDSS-II has been provided by the Alfred P. Sloan Foundation, the Participating Institutions, the National Science Foundation, the U.S. Department of Energy, the National Aeronautics and Space Administration, the Japanese Monbukagakusho, the Max Planck Society, and the Higher Education Funding Council for England. The SDSS Web Site is http://www.sdss.org/.

The SDSS is managed by the Astrophysical Research Consortium for the Participating Institutions. The Participating Institutions are the American Museum of Natural History, Astrophysical Institute Potsdam, University of Basel, University of Cambridge, Case Western Reserve University, University of Chicago, Drexel University, Fermilab, the Institute for Advanced Study, the Japan Participation Group, Johns Hopkins University, the Joint Institute for Nuclear Astrophysics, the Kavli Institute for Particle Astrophysics and Cosmology, the Korean Scientist Group, the Chinese Academy of Sciences (LAMOST), Los Alamos National Laboratory, the Max-Planck-Institute for Astronomy (MPIA), the Max-Planck-Institute for Astrophysics (MPA), New Mexico State University, Ohio State University, University of Pittsburgh, University of Portsmouth, Princeton University, the United States Naval Observatory, and the University of Washington.

This work has partly been based on observations obtained at the Gemini Observatory, which is operated by the
Association of Universities for Research in Astronomy, Inc., under a cooperative agreement
with the NSF on behalf of the Gemini partnership: the National Science Foundation (United
States), the Science and Technology Facilities Council (United Kingdom), the
National Research Council (Canada), CONICYT (Chile), the Australian Research Council
(Australia), Ministério da Ciência e Tecnologia (Brazil) 
and Ministerio de Ciencia, Tecnología e Innovación Productiva  (Argentina).

This work has partly been based in part on data collected at Subaru Telescope, which is operated by the National Astronomical Observatory of Japan.

David Murray is supported by a Science and Technology Facilities Council studentship.

Dr. A.C. Day-Jones is funded by a Fondecyt fellowship.

Dr. Nicholas Lodieu acknowledges funding from the Spanish Ministry of Science and
Innovation through the Ram\'on y Cajal fellowship number 08-303-01-02.

\bibliographystyle{mn2e}
\bibliography{refs}

\clearpage

\end{document}